\documentclass[imslayout,preprint]{imsart}
\pdfoutput=1
\usepackage[left=1.25in,right=1.25in,top=1in,bottom=1in]{geometry}

\usepackage[T1]{fontenc}
\usepackage{ae,aecompl}
\usepackage{graphicx}
\usepackage{amsmath,amssymb,amsthm,amsfonts}

\usepackage{booktabs}
\usepackage{mathrsfs}
\usepackage{url}
\usepackage[table]{xcolor}
\usepackage{natbib}

\startlocaldefs

\def\ind{\ensuremath{\stackrel{\text{\tiny ind}}{\sim}}}
\def\argmin{\mathop{\rm arg\,min}\limits}%
\def\argmax{\mathop{\rm arg\,max}\limits}%
\let\hat\widehat%
\let\tilde\widetilde%
\def\given{{\,|\,}}
\def\Pr{{\ensuremath{\mathbb P}}}%
\def\Exp{{\ensuremath{\mathbb E}}}%
\def\Var{{\ensuremath{\text{\sf Var}}}}%

\def\miss{{\ensuremath{\text{\sf miss}}}}%

\begin{document}
\begin{frontmatter}

\title{A Bayesian Change Point Model for Detecting Land Cover Changes in
{MODIS} Time Series}
\runtitle{Bay. Ch. Point Mod. Land Cover in {MODIS} Data}

\begin{aug}
\author{\fnms{Hunter} \snm{Glanz}\corref{}%
\ead[label=e1]{hglanz@calpoly.edu}},%
\author{\fnms{Xiaoman} \snm{Huang}},%
\author{\fnms{Minhui} \snm{Zheng}},%
\and
\author{\fnms{Luis} \snm{Carvalho}%
\ead[label=e2]{lecarval@math.bu.edu}}

\runauthor{H. Glanz et al.}
\affiliation{California Polytechnic State University and Boston University}
\address{Department of Statistics\\
California Polytechnic State University\\
1 Grand Avenue, Fac. Offices East\\
San Luis Obispo, California, USA 93407\\
\printead{e1}}
\address{Dept.\@ of Math.\@ and Statistics\\
Boston University\\
111 Cummington Mall\\
Boston, Massachusetts, USA 02215\\
\printead{e2}}
\end{aug}

\begin{abstract}
As both a central task in Remote Sensing and a common problem in many other
situations involving time series data, change point detection boasts a
thorough and well-documented history of study. However, the treatment of
missing data and proper exploitation of the structure in multivariate time
series during change point detection remains lacking. Multispectral, high
temporal resolution time series data from NASA's Moderate Resolution Imaging
Spectroradiometer (MODIS) instruments provide an attractive and challenging
context to contribute to the change point detection literature. In an effort
to better monitor change in land cover using MODIS data, we present a novel
approach to identifying periods of time in which regions experience some
conversion-type of land cover change. That is, we propose a method for
parameter estimation and change point detection in the presence of missing
data which capitalizes on the high dimensionality of MODIS data. We test the
quality of our method in a simulation study alongside a contemporary change
point method and apply it in a case study at the Xingu River Basin in the
Amazon. Not only does our method maintain a high accuracy, but can provide
insight into the types of changes occurring via land cover conversion
probabilities. In this way we can better characterize the amount and types of
forest disturbance in our study area in comparison to traditional change point
methods.
\end{abstract}


\end{frontmatter}

\section{Introduction} 
\label{intro}

To enhance and inform Earth system models, timely and accurate monitoring of
land cover must be
maintained~\citep{bonan2002land,ek2003implementation,running1988general,sterling2008comprehensive}.
Additionally, because the land area affected by humans has expanded
rapidly~\citep{ellis08,goldewijk2001estimating,ramankutty1999estimating,sanderson02,vitousek97}
and society depends to a large extent on terrestrial
ecosystems~\citep{foley05}, high quality information regarding changes in land
cover is crucial for modern land-use policy and natural resource management.

Remote sensing instruments onboard various satellite platforms have been
providing repeated observation of the Earth's surface, enabling continuous
mapping and monitoring of land cover change, especially those caused by human
activities. With continuous missions, some instrument series have observations
over the past few decades (e.g., the Landsat series, the Advanced Very High
Resolution Radiometer (AVHRR) series). A unique sensor named the Moderate
Resolution Imaging Spectroradiometer (MODIS), has been in orbit onboard NASA's
Terra and Aqua satellites since the early 2000s. This instrument strikes a
balance between moderate spatial resolution (250--500 meters) and high revisit
capability, providing time series observations for over a decade.
However, a host of issues plagues MODIS data such as measurement errors,
atmospheric contamination, and variable view
geometry and gridding artifacts~\citep{Roy:2000p3565,Huang:2002p305,Tan:2006p7105}, 
and renders change detection a challenging task due to missing and noisy data.

Various change detection techniques were developed using bi-temporal or
multi-temporal imagery for mapping changes including deforestation, forest
mortality, and urban expansion
(see~\citep{Singh:1989p3461,rogan02,coppin04,Lu:2004p559}). As MODIS time
series grow, more studies have focused on better exploitation of the temporal
information in MODIS data for change detection,
e.g.~\citep{Verbesselt:2010p335,Rahman:2013vd,Huang:2014ii}.
However, due to the volume of data and nature of optical remote sensing
(susceptible to cloud and atmospheric contamination), it remains challenging
to pre-process and fully utilize the time series data. Thus, there is great
need of methods that (i) better address missing data; that (ii) explore the
rich structure in the data in their spectral, temporal, and spatial
dimensions; and that (iii) are robust to noise. 

Most existing methods for change detection in the presence of missing data
attempt to impute or estimate missing data first and then proceed to identify
changes~\citep{lunetta1999remote,lunetta06,boriah2010time}.
Estimation can proceed in a number of ways, including, for example, nearest
neighbor interpolation~\citep{ning2012,zhang2012,jerez2010} or linear,
polynomial, or spline interpolation~\citep{junninen2004methods}. Missing
values can be imputed using multiple imputation~\citep{honaker2010missing} or
expectation-maximization (EM)~\citep{Dempster:1977ul} (for a thorough review
of handling missing data in statistical analyses,
see~\citep{little2002statistical}.) However, since missing data are often
handled separately from and prior to change point estimation, the imputation
does not account for possible large changes and so the resulting change
detection can lack statistical power.

In this paper we introduce and assess a novel, off-line change point detection
model that is tailored to the data characteristics of MODIS time series, i.e.
large and structured.
Our key contribution is to characterize change as transitions in \emph{land
cover}: we assume that the region of study is reasonably homogeneous, with a
predominant ``background'' land cover class, and we evaluate change by
implicitly classifying land cover and contrasting estimated classes to the
background. This way, we can not only detect changes but also understand their
nature; for instance, we can better assess if native forest was burned,
logged, or converted to cropland. By exploiting land cover information from
training data, we specify a Bayesian hierarchical model to detect
distributional, conversion-type changes in multispectral time series while
accounting for missing data. In addition, as opposed to at-most-one-change
(AMOC) models that aim at detecting single abrupt disruptions, our formulation
allows for at most two change points and thus also considers possible recovery
from prior disturbances. We describe the change point detection model in
Section~\ref{methods}, and we apply and evaluate our model using a simulation
study (Section~\ref{sec:simstudy}) and a case study
(Section~\ref{sec:casestudy}).

\subsection{Data Description}
\label{data}

To illustrate the main issues that afflict MODIS data, here we describe the
dataset that is used in the case study of Section~\ref{sec:casestudy}. We use the
MODIS 500 meter Nadir BRDF-adjusted Reflectance (NBAR) product, which is
designed to minimize noise due to bidirectional reflectance effects arising
from varying solar and view geometry~\citep{Schaaf:2002p507}. This product
features seven spectral bands designed for land observation, covering visible
to shortwave infrared wavelengths~\citep{MODIS}.

For each pixel in the region of interest and for each year in the
dataset---from 2001 to 2010---we originally obtained time series of 46
NBAR composite values for seven spectral bands. However, for our analysis we
select a temporal subset of 19 observations per year (May to September) in
order to exclude the wet season and reduce the proportion of missing data.
Here it is essential to treat years as the main temporal unit to keep
seasonality effects, including phenology, that characterize land cover
classes. We have verified that this subset still keeps enough seasonality
within the year to distinguish well between classes.

\begin{figure}
\centering
\begin{tabular}{cc}
\includegraphics[width=.47\textwidth]{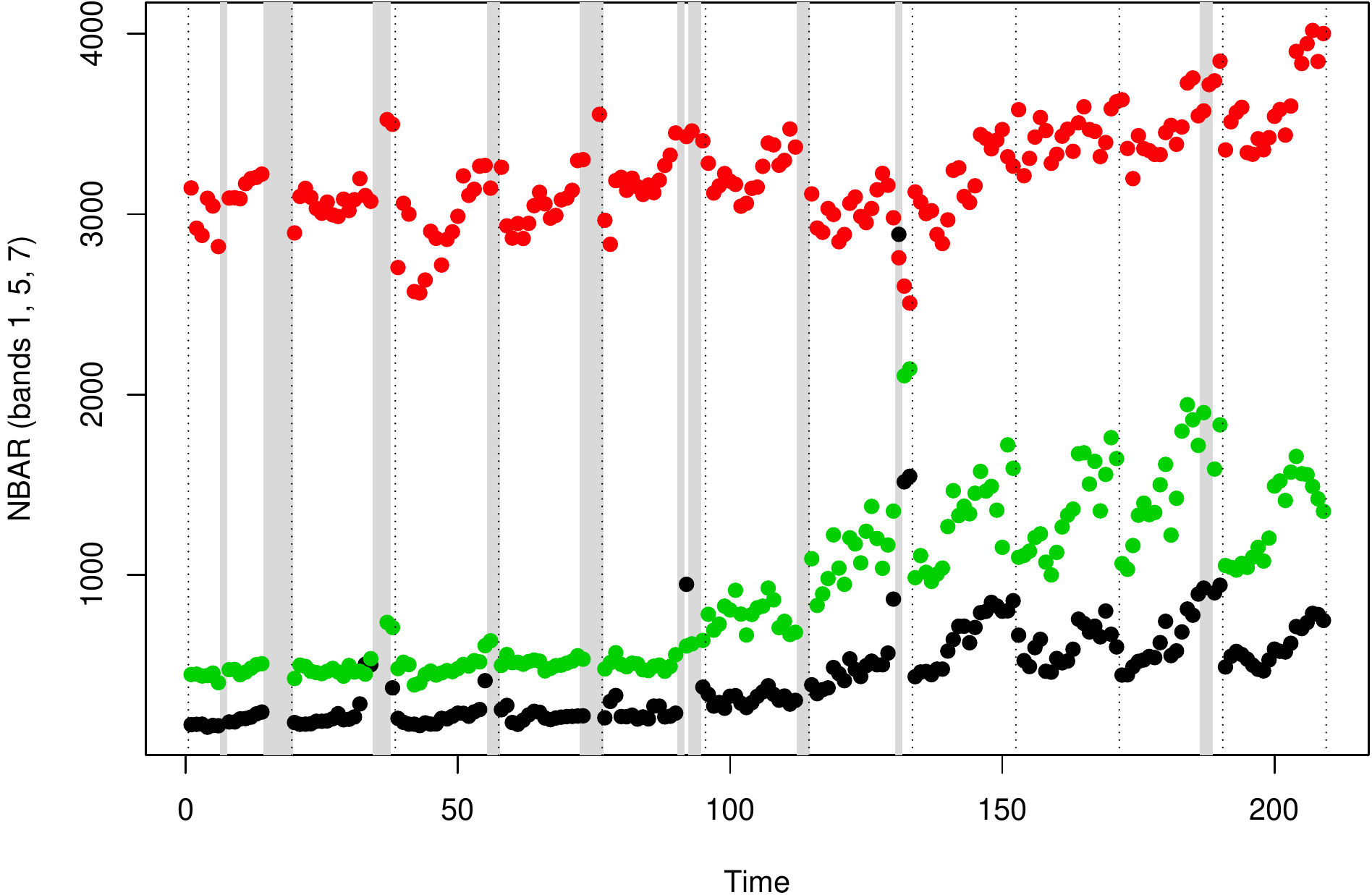} &
\includegraphics[width=.47\textwidth]{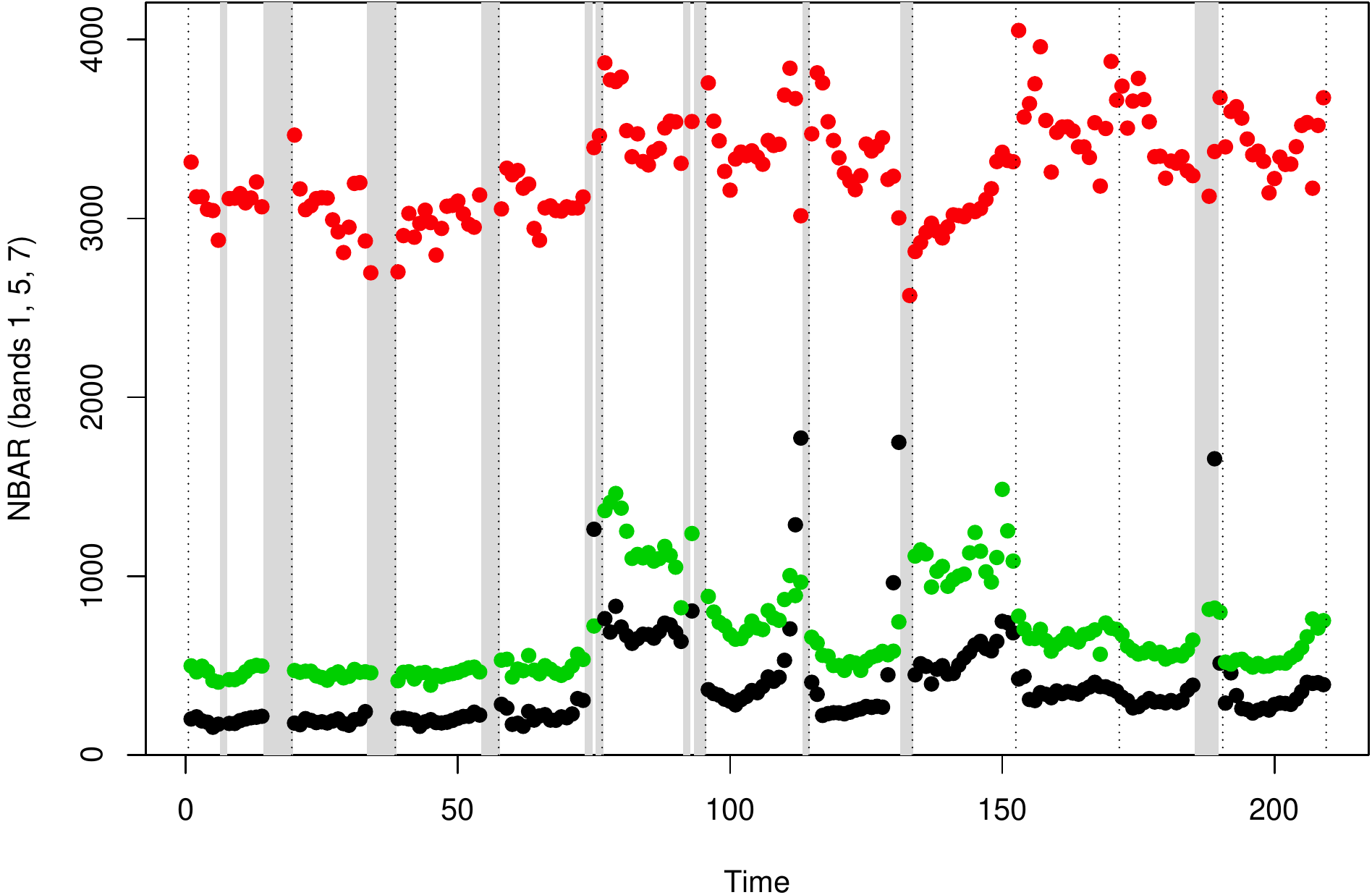}
\end{tabular}
\caption{Spectral-temporal profiles for two representative pixels in a study
area, bands~1 (black), 5 (red), and~7 (green.) Gray horizontal lines mark
missing data in at least one spectral band. Reflectance values have been
multiplied by~10000.}
\label{fig:example}
\end{figure}

As an example, consider the spectro-temporal profiles for two representative
pixels in Figure~\ref{fig:example}. To avoid overcrowding the plot, we only
show three spectral bands (1, 5, and~7.) Gray bands mark missing data
locations in at least one band. As we can see, most years have at least one
time with missing values, and so discarding whole years is unfeasible.
Moreover, since missing data happens more frequently at the end of our annual
time series (i.e. start of wet season), it makes it harder to spot change
between years. Some changes are more evident, as shown in the left plot at
year~6, but some are harder to flag and can be attributed to minor
disturbances, as in the right plot, at year~5. The right plot also highlights
the possibility of recovery: the data profile seems to have returned to its
background land cover state after year~9.

Land cover change detection requires a scheme of land cover classes which
encompasses all major land cover types. We employ a carefully established set
of land cover classes constructed under the International Geosphere-Biosphere
Programme (IGBP)~\citep{igbpdefs}, as defined in Table~\ref{tab:igbpdefs}.

\begin{table}
\caption{Land cover class definitions within the International
Geosphere-Biosphere Programme (IGBP.)}
\label{tab:igbpdefs}
\centering
\footnotesize
\begin{tabular}{cp{.15\textwidth}p{.7\textwidth}} \toprule
\textbf{Class}	&	\textbf{Class name}	&	\textbf{Description}	\\ \midrule
1	&	Evergreen Needleleaf Forests	&	Lands dominated by trees with a percent
canopy cover $>$60$\%$ and height exceeding 2~meters. Almost all tree remain
green all year. Canopy is never without green foliage.	\\
2	&	Evergreen Broadleaf Forests	&	Lands dominated by trees with a percent
canopy cover $>60\%$ and height exceeding 2~meters. Almost all tree remain
green all year. Canopy is never without green foliage.	\\
3	&	Deciduous Needleleaf Forests	&	Lands dominated by trees with a percent
canopy cover $>$60$\%$ and height exceeding 2~meters. Consists of seasonal
needleleaf tree communities with an annual cycle of leaf-on and leaf-off
periods. \\
4	&	Deciduous Broadleaf Forests	&	Lands dominated by trees with a percent
canopy cover $>$60$\%$ and height exceeding 2~meters. Consists of seasonal
broadleaf tree communities with an annual cycle of leaf-on and leaf-off
periods. \\
5	&	Mixed Forests	&	Lands dominated by trees with a percent canopy cover
$>$60$\%$ and height exceeding 2~meters. Consists of tree communities with
interspersed mixtures or mosaics of the other four forest cover types. None of
the forest types exceeds 60$\%$ of landscape. \\
6	&	Closed Shrublands	&	Lands with woody vegetation less than 2~meters tall
and with shrub canopy cover is $>$60$\%$. The shrub foliage can be either
evergreen or deciduous. \\
7	&	Open Shrublands	&	Lands with woody vegetation less than 2~meters tall and
with shrub canopy cover is 10--60$\%$. The shrub foliage can be either
evergreen or deciduous. \\
8	&	Woody Savannas	&	Lands with herbaceous and other understorey systems, and
with forest canopy cover between 30--60$\%$. The forest cover height exceeds
2~meters. \\
9	&	Savannas	&	Lands with herbaceous and other understorey systems, and with
forest canopy cover between 10--30$\%$. The forest cover height exceeds
2~meters. \\
10	&	Grasslands	&	Lands with herbaceous types of cover. Tree and shrub cover
is less than 10$\%$. \\
11	&	Permanent Wetlands	&	Lands with a permanent mixture of water and
herbaceous or woody vegetation that cover extensive areas. The vegetation can
be present in either salt, brackish, or fresh water. \\
12	&	Cropland	&	Lands covered with temporary crops followed by harvest and a
bare soil period (e.g.\ single and multiple cropping systems). Note that
perennial woody crops will be classified as the appropriate forest or shrub
land cover type. \\
13	&	Urban and Built-Up	&	Lands covered by building and other man-made
structures. \\
14	&	Cropland/Nat. Veg. Mosaics	&	Lands with a mosaic of croplands,
forest, shrublands, and grasslands in which no one component comprises more
than 60$\%$ of the landscape. \\
15	&	Snow and Ice	&	Lands under snow and/or ice cover throughout the year. \\
16	&	Barren	&	Lands exposed soil, sand, rocks, or snow and never has more
than 10$\%$ vegetated cover during any time of the year. \\
17	&	Water Bodies	&	Oceans, seas, lakes, reservoirs, and rivers. Can be
either fresh or salt water bodies. \\ \bottomrule
\end{tabular}
\normalsize
\end{table}

\subsection{Prior and Related Work}
\label{priorwork}
Change point detection methods have been applied extensively in various fields
of environmental and climate monitoring, to problems such as rates of Tropical
cyclone activity, precipitation and temperature trends, and fishery population
regime change~\citep{Elsner:2000tr,Chu:2004wn,Rodionov:2005tx,Solow:2005wr}.
Statistically, the general change point problem can be categorized into
on-line (real time)~\citep{Fearnhead:2007p21229} and off-line (retrospective)
frameworks. Additionally, approaches to change point detection typically
involve specifying which types of change to look for. Previous methods for
detecting change vary by the following change types: mean-type
shifts~\citep{Shao:2010gh,Lund:2002ud}, variance
change~\citep{Galeano:2007jk}, or change in
distribution~\citep{Basseville:1993wp,Lee:2010dq,Tsay:1988wu,Song:2007uz,Gombay:2008ir}.
Popular approaches include time series models, sequential testing, special
forms of regression, and Bayesian
techniques~\citep{Menzefricke:1981ud,Booth:1982tp,Stephens:1994ww,Perreault:2000jm,Fearnhead:2006p21221}.

With continuous data collection and growing time series from the MODIS
instruments, many studies in the remote sensing literature have put more
emphasis on exploring temporal information for land cover change detection.
Some of these methods detect change at the pixel level using change indices
derived from annual time
series~\citep[e.g.,][]{Linderman:2005p311,Mildrexler:2009p552,Coops:2009jv}.
Other studies developed temporal trajectory-based change detection algorithms
such as temporal segmentation, structural break test, and distance-metric
based
methods~\citep[e.g.,][]{Verbesselt:2010p335,SullaMenashe:2013cc,Huang:2014ii}.
While some of these methods have demonstrated feasibility for large area
application, it remains challenging to pre-process the data for gap-free input
and reduce spurious detection of change due to noise.  

In this paper, we use the change detection method described
in~\citep{Huang:2014ii} for comparison with our method. It is a distance
metric-based change detection method for identifying changed pixels at annual
time steps using 500 m MODIS NBAR time series data. The approach we describe
uses distance metrics to measure (i) the similarity between a pixel's annual
time series to annual time series for pixels of the same land cover class, and
(ii) the similarity between annual time series from different years at the same
pixel. The combination of two distance metrics used both spatial (regional
land cover related knowledge) and temporal information, and was shown to
compare well with reference information derived from higher spatial resolution
data. A set of essential pre-processing steps, including gap-filling,
smoothing and temporal subsetting of MODIS 500 m NBAR time series, were also
described as part of the approach.

\section{Model and Methods}
\label{methods} 

Consider, for each \emph{year} $i = 1, \ldots, J$, and each \emph{pixel}
$v$ in the region of interest $\mathscr{R}$, the vector observation $X_{iv}$
containing data from $B$ spectral bands and $T$ within-year time points. For
example, in the data described in Section~\ref{data}, $B=7$, $T=19$, and
$J=10$. Since our data contain physical dimensions we exploit these features
by partitioning the variation in the data into spectral and temporal
components. Moreover, we expect land cover classes to have different mean
profiles and different variances so we are able to distinguish them. Thus, if
$\mathscr{C}$ is the set of land cover classes and $W_v \in \mathscr{C}$ codes
for the land cover class of pixel $v$, we start by modeling the data using a
matrix normal distribution~\citep{dawid81}, or, equivalently,
\begin{equation}
\label{eq:lik}
X_{iv} \given W_v = g \ind N(\mu_g, \Sigma_s \otimes \Sigma_{tg}),
\end{equation}
where $\otimes$ denotes the Kronecker product. That is, instead of assuming
that our multivariate normal data have a single $BT \times BT$ covariance
matrix we employ a Kronecker structured covariance matrix which isolates the
spectral covariance in a $B \times B$ matrix, $\Sigma_s$, and the temporal
covariance in a $T \times T$ matrix, $\Sigma_{tg}$. Note that we assume that
spectral variation ($\Sigma_s$) transcends land cover class, and thus only
allow the means ($\mu_g$) and temporal covariances ($\Sigma_{tg}$) to vary
with land cover class $g$. In this way, we reduce the dimensionality of
parameters to be estimated while keeping a parsimonious model
structure~\citep{glanz2014isprs}. In addition, since the temporal profiles
$\mu_g$ capture seasonality and temporal variability is represented in
$\Sigma_{tg}$, we do not need to explicitly model auto-correlation.

The separable nature of the variance also has the advantage of allowing us to
reduce the dimensionality of the data using a focused PCA compression. If
$\Sigma_s = P \text{Diag}(\lambda_{1:B}) P^\top$ is the eigen-decomposition of
$\Sigma_s$, we select the $K < B$ largest eigenvalues and, regarding $X_{iv}$
as a matrix with $B$ rows, we define a compressed version of $X_{iv}$ as
\begin{equation}
\label{eq:pca}
X^*_{iv} := {\text{Diag}(\lambda_{1:K})}^{-1} P_{1:K}^\top X_{iv}.
\end{equation}
This transformation is equivalent to approximating $\Sigma_s$ using $K$
eigenvectors, $\Sigma_s \approx \Sigma_s^* :=
P_{1:K} \text{Diag}(\lambda_{1:K}) P_{1:K}^\top$, and
decorrelating the columns of $X_{iv}$ by $\Sigma_s^*$.

Given the very large size of the data, we opt to learn land cover parameters
$\mu_g$, $\Sigma_s$, and $\Sigma_{tg}$ in a pre-processing step instead of
jointly with change point estimation. To this end, we adopt the EM method
proposed in~\citep{glanz2014isprs} and apply it to an independent training
dataset. This kind of prior elicitation is similar to empirical Bayes
approaches~\citep{empbayes} and aims at simplifying the model and alleviating
the computational burden of inference. To simplify the notation, for the
remainder of this article we denote $\Sigma_g = \Sigma_s \otimes \Sigma_{tg}$.

While~\eqref{eq:lik} gives a parametric model for the annual data at a pixel,
we require a way to detect changes in land cover when these observations
contain missing values. In pursuit of a change point \emph{year} for each
pixel, if it exists, we devise an EM algorithm which accounts for the missing
data present throughout our region of interest. The following section details
our hierarchical model and estimation procedure for identifying a change in
land cover.

\subsection{Change Point Hierarchical Model and Parameter Estimation}
In our scenario, the annual data for each pixel, $X_{iv}$, are assumed to be
conditionally independent of both data in other years at pixel $v$ as well as
data and potential changes in other pixels. To model change, we allow the year
sequence $1, \ldots, J$ to be \emph{segmented} according to $\rho = (\rho_1,
\rho_2)$, $1 \le \rho_1 \le \rho_2 \le J$, such that the segment $\rho_1 + 1,
\ldots, \rho_2$ is in the ``change'' state, and the pre- and post-change
segments $1, \ldots, \rho_1$ and $\rho_2+1, \ldots, J$ are in the
``background'' state. This way, if $\rho_2 < J$ we have recovery from change
to background. Lack of change is represented by $\rho_1 = \rho_2 = J$, the
only case when $\rho_1 = \rho_2$, that is, for any other configuration we have
$\rho_1 < \rho_2$.

For each pixel $v$, we assume the data in the background segment, i.e.\ up to
the change point year $\rho_{1v}$ and after change point year $\rho_{2v}$,
follow a multivariate normal distribution with mean $\mu_{0v}$, and
the data in the change segment, i.e.\ from years $\rho_{1v}+1$ to $\rho_{2v}$,
follow another multivariate normal distribution with mean $\mu_{cv}$.
In addition, to accommodate more flexibility from pixel to pixel,
we add a new level to our model and incorporate land cover class information
via prior distributions for $\mu_{0v}$ and $\mu_{cv}$. Specifically, we set
conjugate priors $\mu_{0v} \sim N(\mu_F, \Sigma_F)$ where $\mu_F$ and
$\Sigma_F$ denote the mean and covariance of our background class, say
Evergreen Broadleaf Forest (EBF); and $\mu_{cv} \given W_v = g \sim
N(\mu_g, \Sigma_g)$, where now $W_v \in \mathscr{C}$ indicates the land cover
class to which pixel $v$ has transitioned in case of a change.
The actual observations $X_{iv}$ now spread around $\mu_{0v}$ and $\mu_{cv}$
according to variance scales $\kappa_0$ and $\kappa_c$:
\begin{equation}
\label{eq:likcp}
X_{iv} \given \mu_{0v}, \mu_{cv}, \rho_v \ind
I(i \in \text{BG}(\rho_v)) N(\mu_{0v}, \kappa_0 I_{BT})
+ I(i \not\in \text{BG}(\rho_v)) N(\mu_{cv}, \kappa_c I_{BT}),
\end{equation}
where $I(\cdot)$ is the indicator function, the background
segment of $\rho_v$ is
$\text{BG}(\rho_v) = \{i : i \le \rho_{1v} \text{~or~} i > \rho_{2v}\}$, and
thus change positions $i \not\in \text{BG}(\rho_v)$ correspond to
$\rho_{1v} < i \le \rho_{2v}$.
Since the change affects the mean
yearly temporal profiles $\mu_{0v}$ and $\mu_{cv}$, we can regard them as
smoothed versions of $X_{iv}$ and so this hierarchical model is similar in
spirit to the smoothing approach of~\citet{lunetta06}.
However, since our interest does not lie in the mean profile parameters
$\mu_{0v}$ and $\mu_{cv}$, we can further simplify our model by marginalizing
them out to obtain:
\begin{equation*}
X_{iv} \given \rho_v, W_v = g \ind
I(i \in \text{BG}(\rho_v)) N(\mu_F, \Sigma_F + \kappa_0 I_{BT})
+ I(i \not\in \text{BG}(\rho_v)) N(\mu_g, \Sigma_g + \kappa_c I_{BT}).
\end{equation*}

\begin{figure}
\centering
\begin{tabular}{cc}
\includegraphics[width=.47\textwidth]{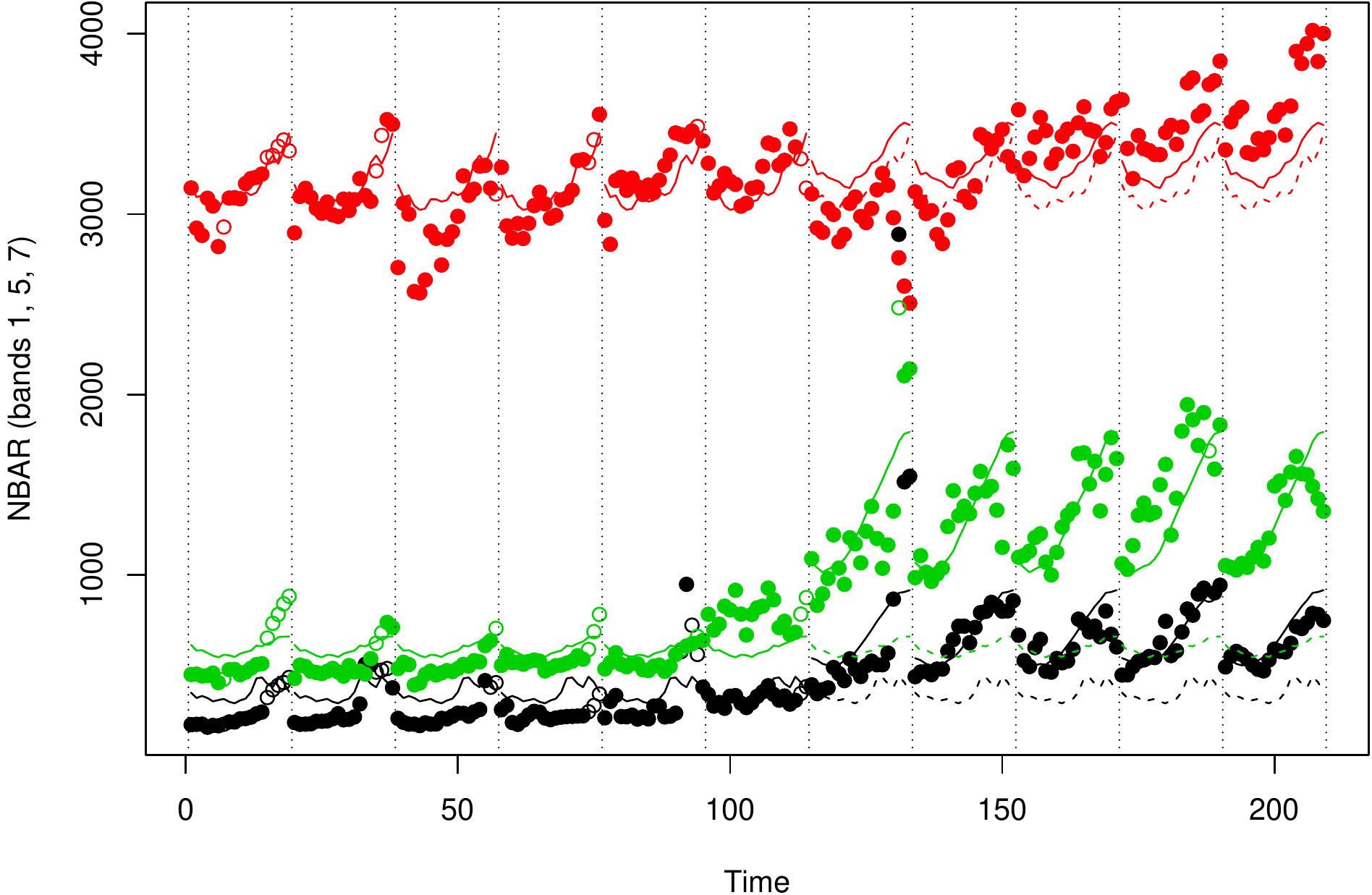} &
\includegraphics[width=.47\textwidth]{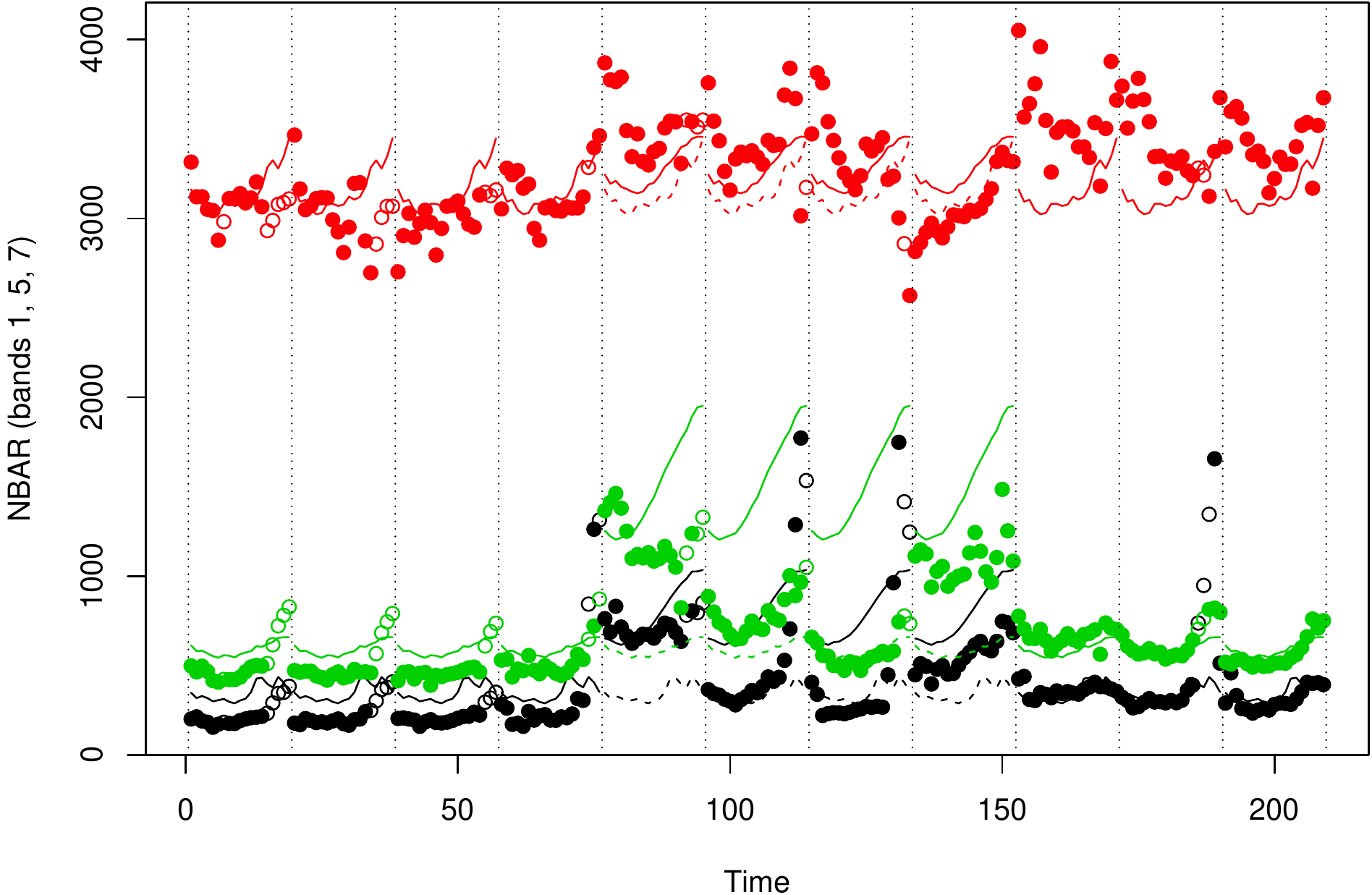}
\end{tabular}
\caption{Spectral-temporal profiles for two representative pixels in study
area along with estimated mean profiles for background and change land cover
classes. Hollow points mark EM-imputed values. Dashed lines during change
periods represent mean profiles under background land cover class for
comparison. Reflectance values have been multiplied by~10000.}
\label{fig:example-fitted}
\end{figure}

As an example, Figure~\ref{fig:example-fitted} depicts $X_{iv}$ for the two
representative pixels that were shown in Figure~\ref{fig:example}, along with
estimated $\hat{\mu}_{0v}$, $\hat{\mu}_{cv}$, and $\hat{\rho}_v$ using the
EM~method described in Section~\ref{sec:cpdem}. For the pixel on the left
panel, $\hat{\rho}_{1v} = 6$ and $\hat{\rho}_{2v} = 11$ (no recovery), while
for the pixel on the right panel we have $\hat{\rho}_{1v} = 4$ and
$\hat{\rho}_{2v} = 8$.

To set a weakly informative prior on $\rho_v$, we settle on a hierarchy that 
depends on two probabilities---the probability of a change occurring, $\pi_0$,
and, given that a change occurred, the probability of recovery $\pi_R$---and
we specify that configurations with the same number of change points are
equally likely. Thus, the probabilities of no change, change without recovery
(one change point), and change with recovery (two change points) are given,
respectively, by
\begin{equation}
\label{eq:cpprior}
\begin{split}
\Pr(\rho_{1v} = \rho_{2v} = J) &= 1 - \pi_0, \\
\Pr(\rho_{1v} < \rho_{2v} = J) &= \frac{\pi_0(1 - \pi_R)}{J - 1},~\text{and} \\
\Pr(\rho_{1v} < \rho_{2v} < J) &= \frac{\pi_0\pi_R}{\binom{J - 1}{2}}. \\
\end{split}
\end{equation}
Finally, we set
$W_v \given \boldsymbol{\alpha} \sim \text{\sf MN}(1, \boldsymbol{\alpha})$ to
depend on a region-wise parameter $\boldsymbol{\alpha}$ that tells the \emph{a
priori} probability of changing to a certain class in $\mathscr{C}$, and
elect a conjugate prior
$\boldsymbol{\alpha} \sim \text{\sf Dir}(\boldsymbol{\pi})$.
The specification of $\boldsymbol{\pi}$ provides an advantageous flexibility
that we can exploit to inform the model of land cover classes we anticipate
seeing after a change has occurred, making our approach particularly well
suited for changes in the form of land cover conversions.

Our model can accommodate changes in mean or covariance and benefits from a
Bayesian approach which incorporates potential \emph{a priori} information
about existence and location of a change point. Our ultimate goal with this
model consists of inferring the change point locations $\rho_v$ for every
pixel in the region of interest, a task we discuss next.

\subsection{Identifying Change Points via Expectation-Maximization}
\label{sec:cpdem}
To account for missing data, we partition the data in year $i$ and pixel $v$
as $X_{iv} = (Y_{iv}, Z_{iv})$ where $Y_{iv}$ are actual observed data and
$Z_{iv}$ are missing values. The missing entries $Z_{iv}$ can occur at
multiple times within year $i$ and at multiple spectral bands, and these
entries can vary from pixel to pixel. We assume that $Z_{iv}$ occur missing at
random and represent them as $Z = {\{Z_{iv}\}}_{i=1,\ldots,T, v \in
\mathscr{R}}$ the whole collection of missing values in the dataset (and
similarly for $Y = \{Y_{iv}\}$, the observed values.)

To estimate our parameters of interest
$\Theta = \{{\{\rho_v\}}_{v \in \mathscr{R}}, \boldsymbol{\alpha}\}$
we select a representative of the posterior distribution
$\Pr(\Theta \given Y)$ such as the maximum \emph{a posteriori} (MAP) estimator
\begin{equation}
\label{eq:map}
\hat{\Theta}
= \argmax_{\Theta} \sum_{W} \int
\Pr(\Theta, Z, W \given Y) \,\text{d}Z
= \argmax_{\Theta} \Pr(\Theta \given Y),
\end{equation}
where $W = {\{W_v\}}_{v \in \mathscr{R}\,:\,\rho_{1v} < J}$; that is, we
marginalize the nuisance parameters $Z_v$, the missing values, and the change
land class $W_v$ across all pixels $v \in \mathscr{R}$.
While a traditional Bayesian approach relies on estimating
$\Pr(\Theta \given Y)$ using Markov chain Monte Carlo (MCMC)
methods~\citep{robertcasella99,bda}, here we adopt an EM routine for
computational expediency since we anticipate assessing change in large
datasets that often comprise millions of pixels. Under this setup, we regard
both $Z$ and $W$ as latent variables and wish to estimate directly the MAP
in~\eqref{eq:map} by following a procedure that starts at some arbitrary
$\Theta^{(0)}$ and iteratively updates
\begin{equation}
\label{eq:emupdate}
\begin{split}
\Theta^{(t+1)} &= \argmax_{\Theta} Q(\Theta, \Theta^{(t)})
:= \argmax_{\Theta} \Exp_{Z, W \given Y; \Theta^{(t)}}
\big[ \log \Pr(\Theta, Z, W, Y) \big] \\
&= \argmax_{\Theta} \Exp_{Z, W \given Y; \Theta^{(t)}}
\big[ \log \Pr(\Theta, Z, W \given Y) \big]
\end{split}
\end{equation}
until convergence. Function $Q$ computes the expectation (E) step, while the
update in~\eqref{eq:emupdate} performs the maximization (M) step.

In the spirit of a cyclic gradient descent approach, we alternate between
updating the ``global'' parameter $\boldsymbol{\alpha}$ and then updating
change points $\rho_v$ for each pixel $v$. This procedure is similar to a
block version of an expectation conditional maximization (ECM)
routine~\citep{mengrubin93}. The details are as follows:
\begin{enumerate}
\item Start at arbitrary $\Theta^{(0)}$; for example, set
$\alpha^{(0)}_k = \pi_k / \sum_{g \in \mathscr{C}} \pi_g$,
for $k \in \mathscr{C}$, and $\rho^{(0)}_{1v} = \rho^{(0)}_{2v} = J$ for all
pixels $v \in \mathscr{R}$.

\item For $t = 1, 2, \ldots$ (until convergence) do
  \begin{enumerate}
  \item For $k \in \mathscr{C}$ do: update
  \begin{equation}
  \label{eq:alphaem}
	\alpha_k^{(t+1)} =
  \frac{\sum_{v\,:\,\rho_{1v}^{(t)} < J} \Pr(W_v = k \given Y_v; \Theta^{(t)})
    + \pi_k - 1}
  {N_v^{(t)} + \sum_{g \in \mathscr{C}} \pi_g - |\mathscr{C}|},
  \end{equation}
  where $N_v^{(t)} = |\{v\,:\,\rho_{1v}^{(t)} < J\}|$ is the number of pixels
  with changes and
  \begin{equation}
  \label{eq:wvem}
	\Pr(W_v = k \given Y_v; \Theta^{(t)}) =
  \frac{\alpha_k^{(t)} \Pr(Y_v \given W_v = k; \Theta^{(t)})}
  {\sum_{g \in \mathscr{C}} \alpha_g^{(t)}
  \Pr(Y_v \given W_v = g; \Theta^{(t)})}.
  \end{equation}

  We note that if we denote by $\miss(X)$ and $-\miss(X)$ the indices of
  missing and non-missing values in $X$ respectively then
  \[
  Y_{iv} \given W_v = k \ind
  N(\mu_{k, -\miss(X_{iv})}, \Sigma_{g, -\miss(X_{iv}), -\miss(X_{iv})}),
  \]
  which we can use to compute $\Pr(Y_v \given W_v = k; \Theta^{(t)})$
  in~\eqref{eq:wvem}.

  \item For each pixel $v$ in the region of interest do: update $\rho_v$ by
  selecting
  \begin{multline}
  \label{eq:rhoem}
  \rho_v^{(t+1)} = \argmin_{\rho} \Bigg\{
  \sum_{i \in \text{BG}(\rho)} S(X_{iv}; \mu_F, \tilde{\Sigma}_F) \\
  + \sum_{i \not\in \text{BG}(\rho)} \sum_{g \in \mathscr{C}}
  \Pr(W_v = k \given Y_v; \Theta^{(t)}) S(X_{iv}; \mu_g, \tilde{\Sigma}_g) \\
  - 2 I(\rho_1 < J) \sum_{g \in \mathscr{C}}
  \Pr(W_v = k \given Y_v; \Theta^{(t)}) \log \alpha_g^{(t+1)}
  - 2 \log \Pr(\rho) \Bigg\},
  \end{multline}
  where
  \begin{multline}
  \label{eq:Sem}
  S(X; \mu_g, \tilde{\Sigma}_g) := \log|\tilde{\Sigma}_g| +
  {(\tilde{X}_g - \mu_g)}^\top \tilde{\Sigma}_g^{-1} (\tilde{X}_g - \mu_g) \\
  + \sum_{j, k \in \miss(X)}
  {({\tilde{\Sigma}_g}^{-1})}_{jk} {(V(X; \tilde{\Sigma}_g))}_{jk}
  \end{multline}
  with $\tilde{\Sigma}_g := \Sigma_F + \kappa_0 I_{BT}$ if $g = F$ and
  $\tilde{\Sigma}_g := \Sigma_g + \kappa_c I_{BT}$ for $g \in \mathscr{C}$.
  More details about the EM-related variables $\tilde{X}_g$, an EM-imputed
  version of $X$, and $V(X; \tilde{\Sigma}_g)$, the conditional variance of
  $X_{\miss(X)}$ given $X_{-\miss(X)}$, can be found in the Appendix.

  The update in~\eqref{eq:rhoem} proceeds by first computing the sufficient
  statistics in~\eqref{eq:Sem} for every $X_{iv}$ and $g = F$ and $g \in
  \mathscr{C}$ and then systematically spanning the possible values of $\rho$
  by including and excluding each year from the background while keeping track
  of the optimal minimum value of the objective in~\eqref{eq:rhoem}.

  \end{enumerate}
\end{enumerate}

We assess convergence by checking if the change in $Q$ between successive
iterations is not significant, that is, we set a threshold $\epsilon$, say
$\epsilon = 10^{-6}$, and stop when
$|Q(\Theta^{(t+1)}, \Theta^{(t)}) - Q(\Theta^{(t)}, \Theta^{(t-1)})| <
\epsilon$. Details on the variables in~\eqref{eq:Sem} and derivations of the
update equations above can be found in the Appendix. However, we can already
notice that inferring the change point locations $\rho_v$ does not involve only
imputation of the missing values, as the quadratic term with $\tilde{X}_g$
implies; we still need to account for the extra variability that arises from
the uncertainty in the missing values, as captured by the term with $V_g(X)$.

In Figure~\ref{fig:example-fitted} we show the results of the proposed method
in two pixels. In both plots, the hollow points are the EM-imputed values
$\tilde{X}_{iv}$, while the mean profile during change, that is, for years
between $\hat{\rho}_{1v} + 1$ and $\hat{\rho}_{2v}$, is taken as
$\hat{\mu}_{cv} = \mu_{g^*}$ with $g^* = \argmax_{k \in \mathscr{C}} \Pr(W_v =
k \given Y_v; \hat{\rho}_v, \hat{\boldsymbol{\alpha}})$ the modal land cover
class. Both pixels belong to the region studied in the next section, where we
provide more details about model fit and inference.
\texttt{R} code implementing this EM routine is available in the Supplementary
Material. 


\section{Data Analysis and Results}
In this section we apply the EM routine from Section~\ref{sec:cpdem} in a
simulation study and a case study involving data from the Xingu River Basin in
the Amazon.

\subsection{Simulation Study}
\label{sec:simstudy}

For the model and EM routine described above, we need to estimate the parameters
of~\eqref{eq:lik} for each of the land cover classes prevalent in our region
of interest: the Xingu River Basin in the southeastern part of the Amazon. We
characterize the regional land cover classes using a set of training sites in
South America located in the Olson ``Tropical and Subtropical Moist Broadleaf
Forests'' biome between 0 and 20$^\circ$S~\citep{Friedl:2010p480}.
Evergreen Broadleaf Forests (class 2) constitute our background (pre- and
post-change) data. 

Our change point simulation study uses a separate set of training sites to
simulate datasets consisting of some pixels with a change and some
without. That is, we construct new, artificial time series profiles based on
an independent collection of 100 pixels which contain different types of
user-identified changes. 

A constructed \emph{no-change} pixel consists of whole years of data being
sampled one year at a time from the portion of these 100 pixels identified as
``background.'' A constructed \emph{change} pixel begins with a randomly
generated change point configuration which partitions the time series into
``background'' and ``change'' periods; then data for these periods
are sampled again, one year at a time, from the ``background'' and ``change''
portions of the 100 training pixels. 
A single \emph{replication} involves simulating 60 no-change pixels and 60
change pixels. For each pixel we stitch together 11 years of data. Each annual
profile consists of data for bands 1 through 7 over 19 time points, as
described in Section~\ref{methods}. A single \emph{batch} consists of 100 such
replications. To explore the influence of missing data we created data for
four batches, and induced minimum proportions of missing data of 20\%, 30\%,
40\% and 50\% in each batch respectively. As a basis for comparison, we
applied our proposed change point method as well as another contemporary
method~\citep{Huang:2014ii} to these simulated data.

\begin{figure}
\begin{center}
	\includegraphics[width=.65\textwidth]{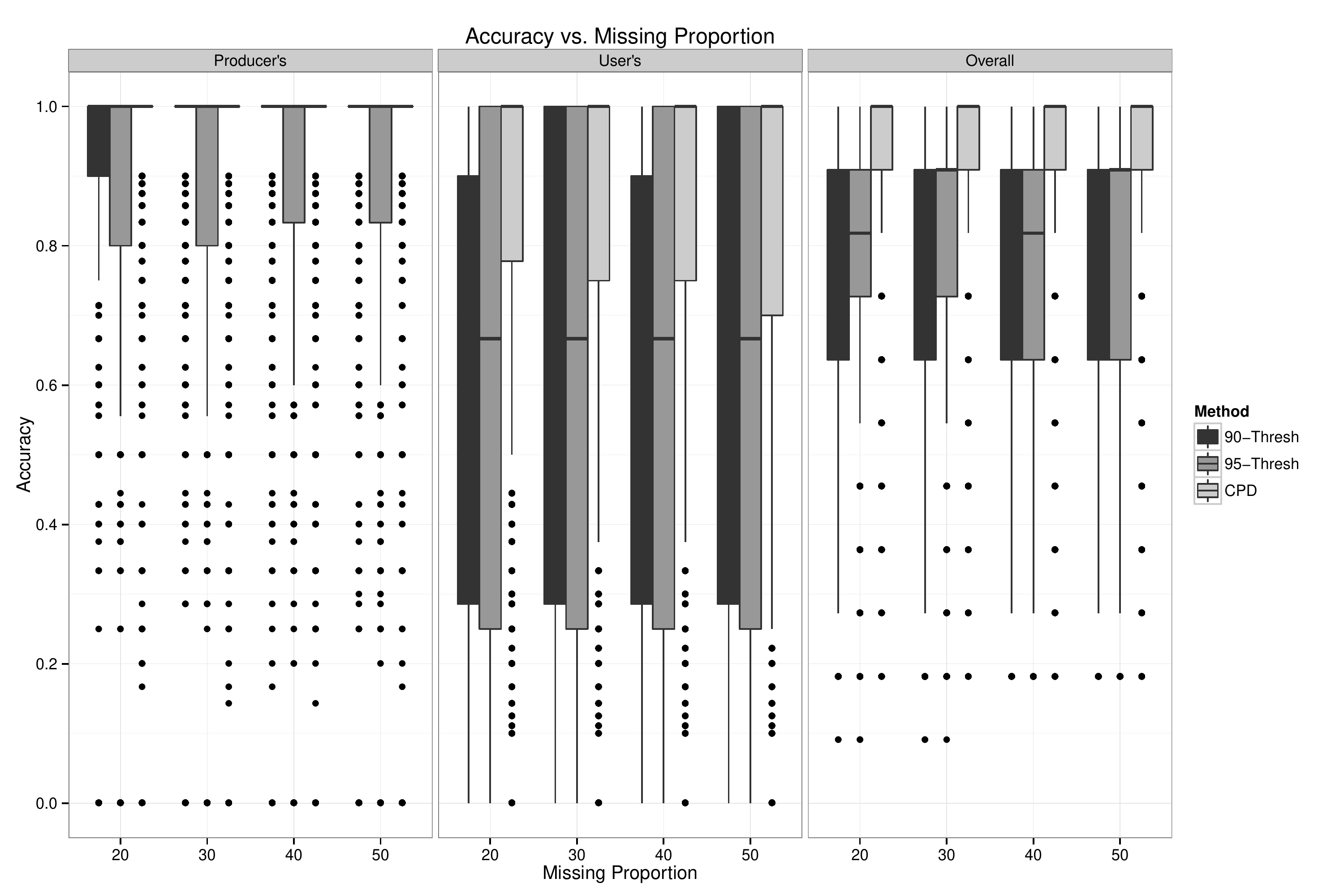}
\end{center}
\caption{Batch accuracies for three different methods applied to simulated
change and no-change data using the metrics outlined in~\eqref{eq:accs}. The
``90-Thresh'' and ``95-Thresh'' correspond to the method
in~\citep{Huang:2014ii} with thresholds of~90\% and~95\%; ``CPD'' corresponds
to our proposed method.}
\label{fig:simstudyaccs}
\end{figure}

\begin{table}
\caption{Average \emph{overall} accuracies in each batch, for each of the three methods.}
\label{tab:accs}
\begin{center}
	\begin{tabular}{cccc} \toprule
		Missing \% & 90-Thresh & 95-Thresh & CPD \\ \midrule
		20 				 & 0.781 & 0.794 & \cellcolor{gray!25}0.920 \\
		30		 		 & 0.782 & 0.796 & \cellcolor{gray!25}0.916 \\
		40 				 & 0.780 & 0.793 & \cellcolor{gray!25}0.913 \\
		50 				 & 0.779 & 0.792 & \cellcolor{gray!25}0.909 \\ \bottomrule
	\end{tabular}
\end{center}
\end{table}

To measure the performance of a change point method we consider three metrics:
producer's accuracy $P$ (sensitivity, recall), user's accuracy $U$ (positive
predictive value, precision), and (overall) accuracy $A$. Given two change
point configurations $\rho$, as classified by the method, and $\tilde{\rho}$,
the ground truth configuration, each metric is given by:
\begin{equation}
\label{eq:accs}
\begin{split}
P(\rho, \tilde{\rho}) &=
\frac{\sum_{i=1}^J I(i \not\in \text{BG}(\rho))
I(i \not\in \text{BG}(\tilde{\rho}))}%
{\sum_{i=1}^J I(i \not\in \text{BG}(\tilde{\rho}))}, \\
U(\rho, \tilde{\rho}) &=
\frac{\sum_{i=1}^J I(i \not\in \text{BG}(\rho))
I(i \not\in \text{BG}(\tilde{\rho}))}%
{\sum_{i=1}^J I(i \not\in \text{BG}(\rho))},~\text{and} \\
A(\rho, \tilde{\rho}) &=
\frac{1}{J} \sum_{i=1}^J
I(i \not\in \text{BG}(\rho)) I(i \not\in \text{BG}(\tilde{\rho})) +
I(i \in \text{BG}(\rho)) I(i \in \text{BG}(\tilde{\rho})).
\end{split}
\end{equation}
If the denominator in either $P$ or $U$ is zero we arbitrarily set them to
zero. The boxplots in Figure~\ref{fig:simstudyaccs} and values in
Table~\ref{tab:accs} summarize the three accuracies mentioned above for our
proposed method as well as the method in~\citep{Huang:2014ii} with thresholds
of 90\% and 95\%. In every situation our proposed method out-performs the
contemporary method at both 90\% and 95\% thresholds. Furthermore, our method
consistently achieves high accuracies ($>$90\%) across substantial amounts of
missing data. The noticeable dip in user's accuracy (as compared with
producer's and overall) across all methods stems from a tendency to identify
an excessively long change period. To adapt to this we could consider updating
our belief about the probability of recovery. After successfully applying our
method to simulated data, we proceed to detect change in a particular region
of the Xingu River Basin.

\subsection{Case Study}
\label{sec:casestudy} 

We apply the EM algorithm described in Section~\ref{sec:cpdem} to an area
(2500 MODIS pixels, $\approx$134 $km^{2}$) in the Xingu River Basin, located
in the Southeastern part of the Amazon in the State of Mato Grosso, Brazil.
The study region has several distinct types of natural vegetation including
moist tropical rainforest, cerrado, and deciduous forest. Despite containing
substantial area of protected indigenous lands, large areas of the basin's EBF
have been converted to agricultural lands for soybean production and cattle
ranching since 2000~\citep{Huang:2014ii}.

To avoid spurious results, we do not consider IGBP classes that are not native
to the study area: 1 (evergreen needleleaf forests), 3 (deciduous needleleaf
forests), and 4 (deciduous broadleaf forest), 11 (permanent wetlands), 13
(urban and built-up), and 15 (snow and ice). Thus, only IGBP classes~5~(MXF),
6~(CSH), 7~(OSH), 8~(WSA), 9~(SAV), 10~(GRA), 12~(CRL), 14~(CRM), 16~(BAR),
and 17~(WAT) are assumed as possible change classes, while IGBP class 2, EBF,
is taken as the background class. For the analysis we assumed that
$\pi_0 = 10^{-10}$, $\pi_R = 0.01$, and that $\kappa_0 = \kappa_c = 5 \cdot
10^4$ which is roughly $1/5$ of the data variance in the classes. The very
stringent value for the probability of change $\pi_0$ aims at providing a more
robust change point inference against outliers. As the probability of recovery
$\pi_R$ suggests, we expect that \emph{a priori} approximately $1\%$ of the
changed pixels actually recover.

To assess our results, we used a high quality Landsat-based deforestation
dataset called PRODES (Monitoring the Brazilian Amazon Gross Deforestation),
produced by Brazil's National Institute for Space Research
(INPE)~\citep{INPE:2012}.
We derived annual sub-pixel fractions of deforestation and the year of change
at MODIS spatial resolution (see~\citep{Huang:2014ii} for details).
In particular, to evaluate the performance of our method, for each pixel
$v \in \mathscr{R}$ we compare the estimated change segmentation given by
$\rho_v$ to reference deforestation percentages $f_v$ using a measure of
\emph{concordance} $C$:
\begin{equation}
\label{eq:concordance}
C(\rho_v, f_v) := \frac{1}{J} \sum_{i=1}^J
I(i \in \text{BG}(\rho_v))(1 - f_{iv}) +
I(i \not\in \text{BG}(\rho_v)) f_{iv}.
\end{equation}
We note that this measure can be seen as an expected accuracy if we regard
$f_{iv}$ as the probability of the $i$-th reference year not being in the
background state.

\begin{figure}
\centering
\includegraphics[width=.95\textwidth]{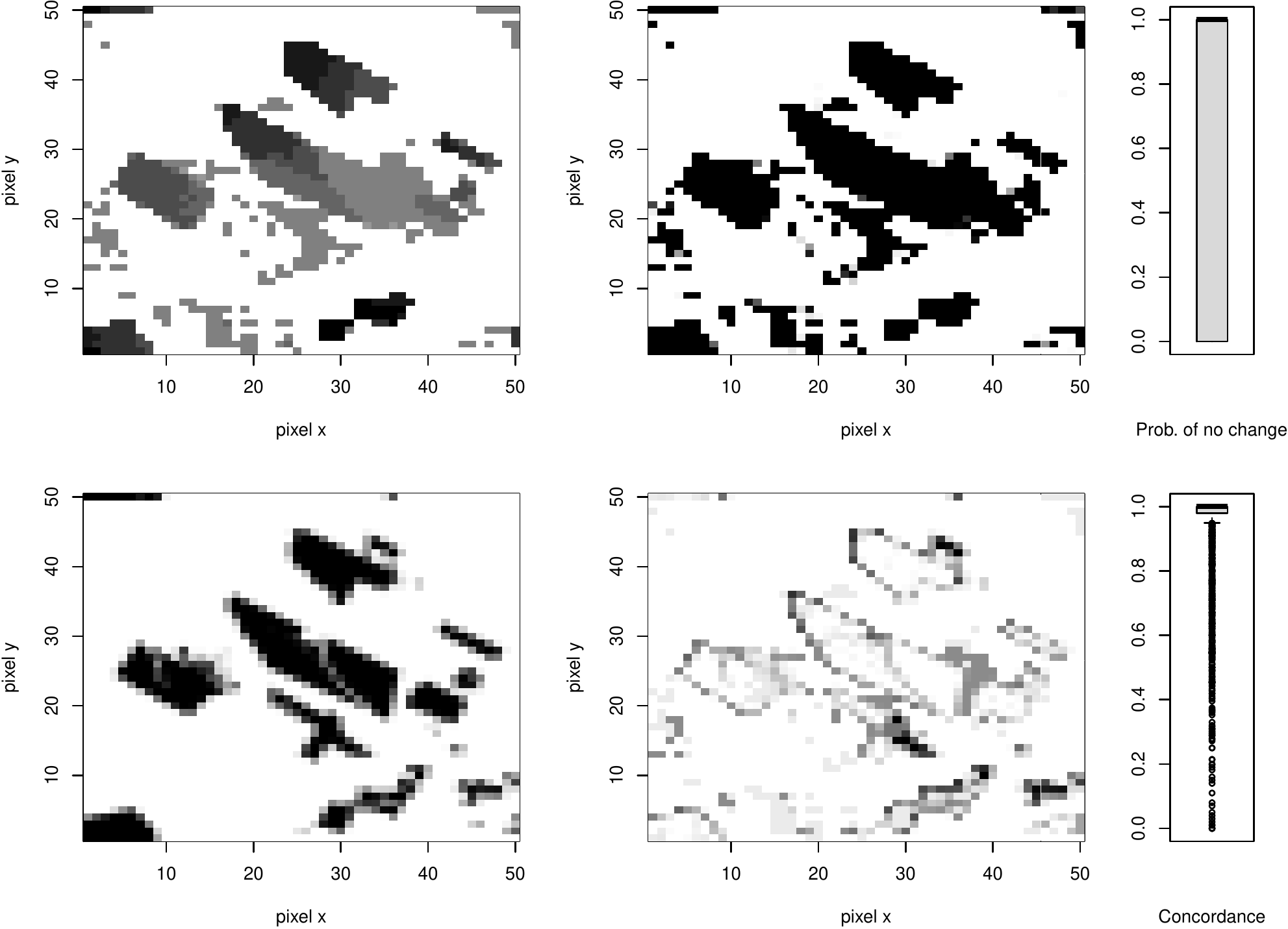}
\caption{Results of change point analysis in the Xingu River study region.
Top row: estimated change points $\hat{\rho}_{1v}$ in the leftmost panel
(darker shades mean earlier changes), conditional probabilities of no change
in the rightmost panels (darker shades represent smaller probabilities.)
Bottom row, left to right: ground-truth reference (darker shades code for
higher deforestation), concordance with estimated change point configurations
(darker shades capture lower concordance), and distribution of concordance
values across pixels.}
\label{fig:xingu-50}
\end{figure}

Figure~\ref{fig:xingu-50} summarizes the inferred changes. In the top left
panel we plot the estimated change year for each pixel $\hat{\rho}_{1v}$ at
the end of the EM procedure for pixel $v$. Darker grays represent earlier
changes and white, in particular, codes for $\hat{\rho}_{1v} = J$, i.e.\@ no
change. The two top rightmost panels show the conditional probability of no
change, that is, $\Pr(\rho_{1v} = J \given Y_v; \hat{\Theta})$, with darker
shades representing smaller probabilities; as can be seen from the contrast in
the spatial pattern and the boxplot, the changes are very accentuated within
clusters. The bottom panels show that the inferred change points are in very
good agreement with the ground-truth reference: the leftmost panel plots
$\max_{i=1,\ldots,J} f_{iv}$, with darker shades representing higher levels of
deforestation; the middle panel plots the concordance measure
in~\eqref{eq:concordance}, darker shades coding for lower concordance values
to highlight contrasts; and the rightmost panel illustrating the distribution
of concordance values across pixels. As we can see, concordance is overall
high and the low values are concentrated either at the borders of change
clusters or at small change ``islands'' (clusters.)

\begin{figure}
\centering
\begin{tabular}{cc}
\includegraphics[width=.47\textwidth]{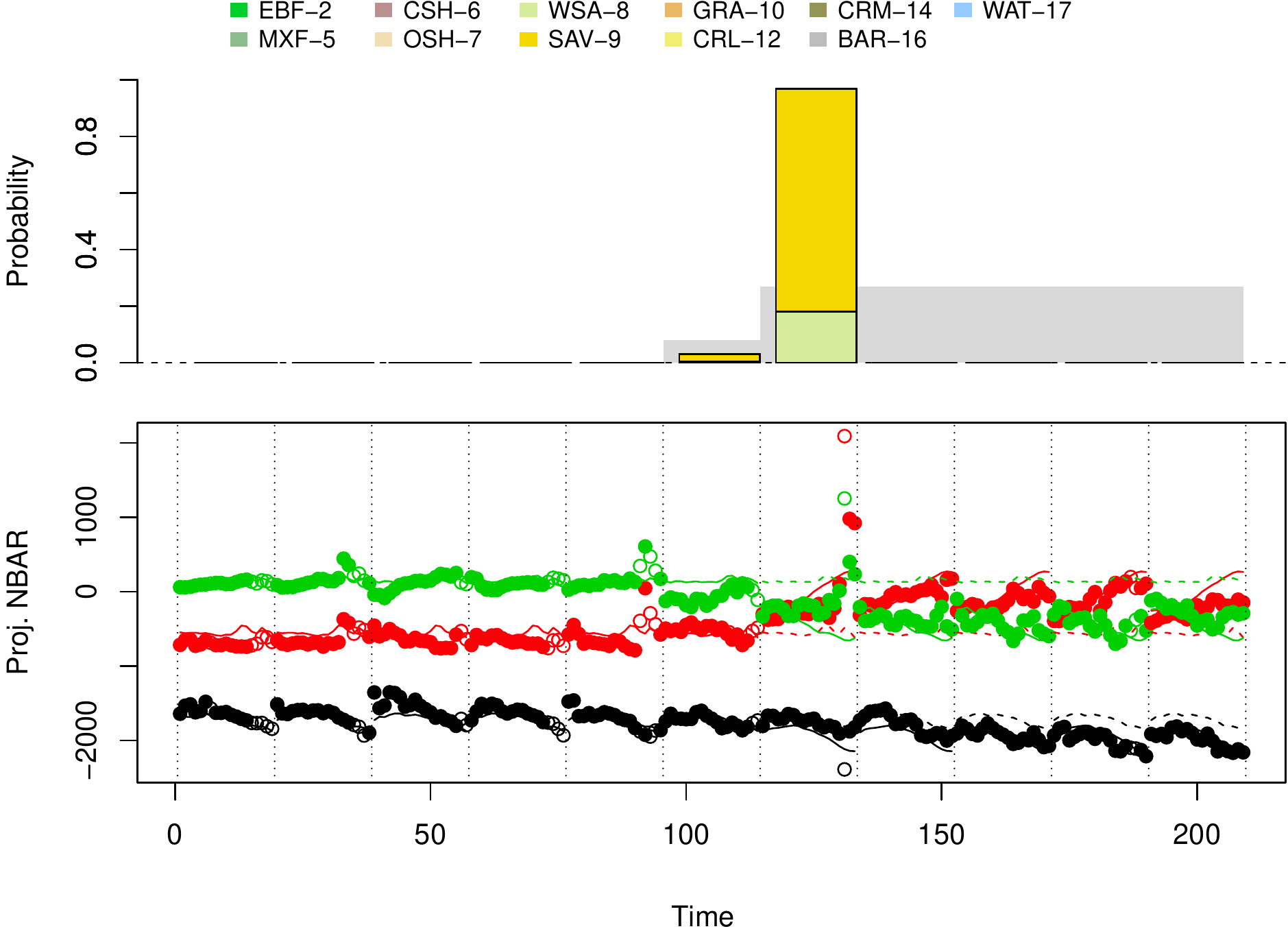} & 
\includegraphics[width=.47\textwidth]{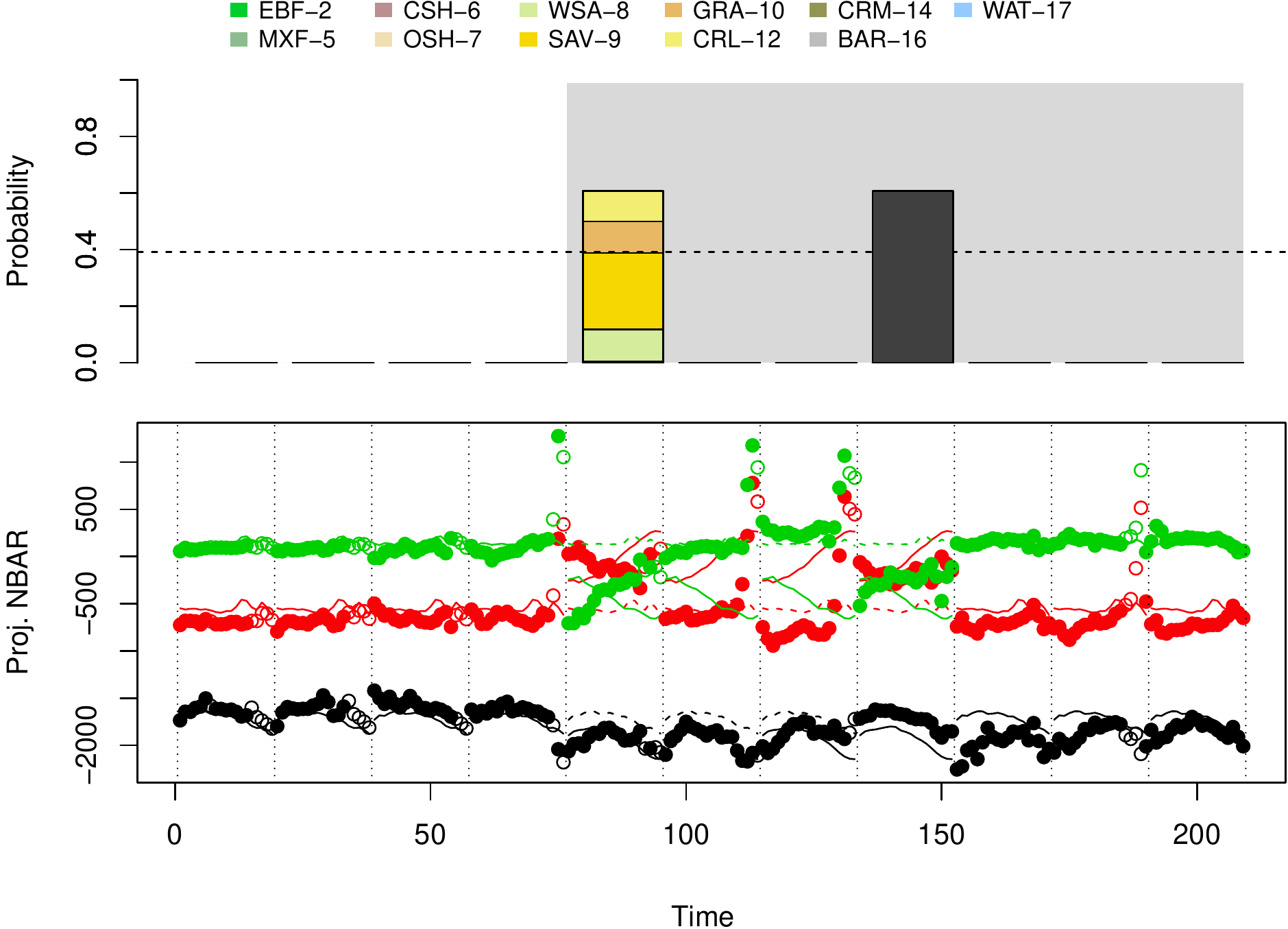} \\
\includegraphics[width=.47\textwidth]{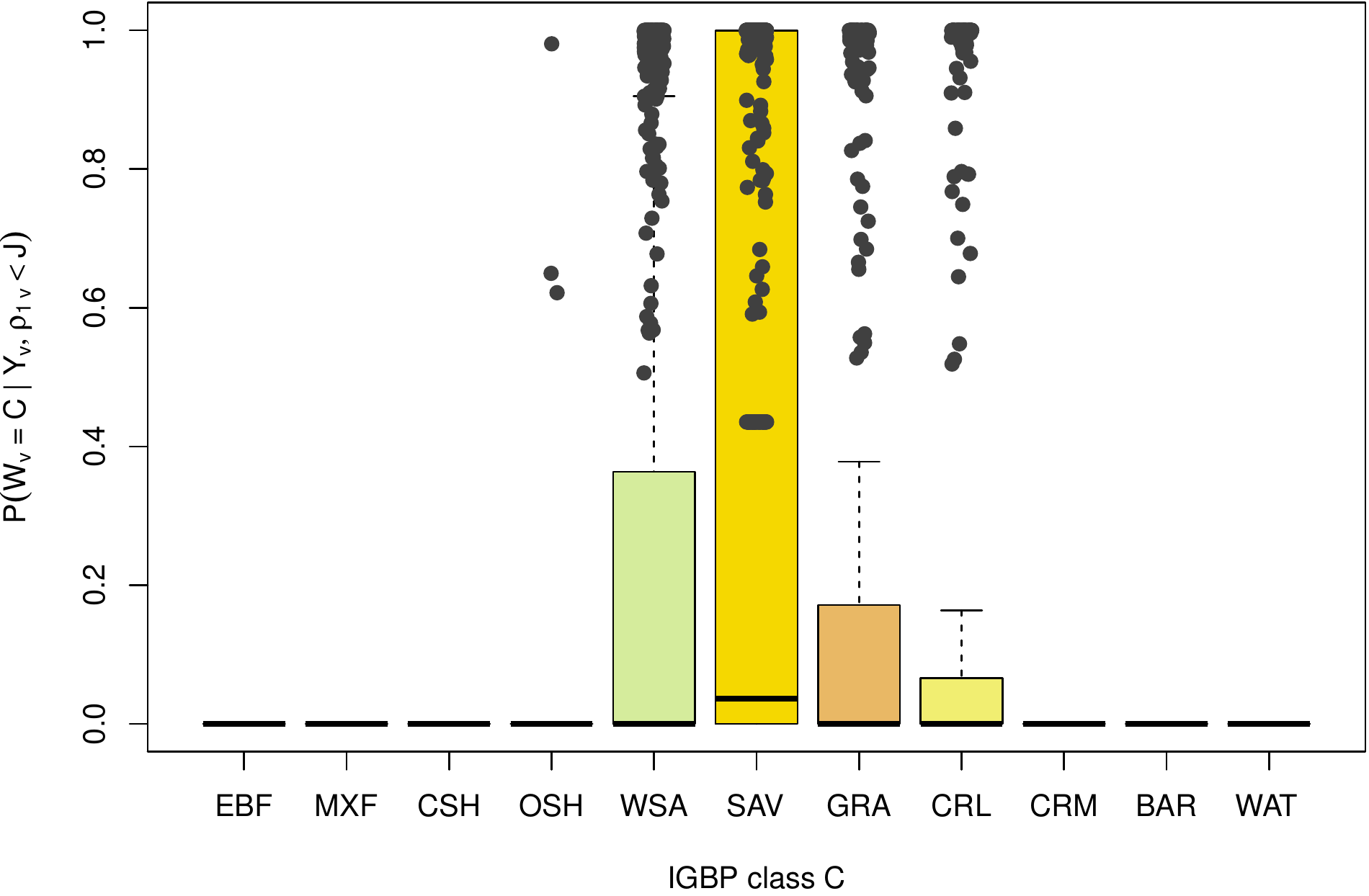} &
\includegraphics[width=.47\textwidth]{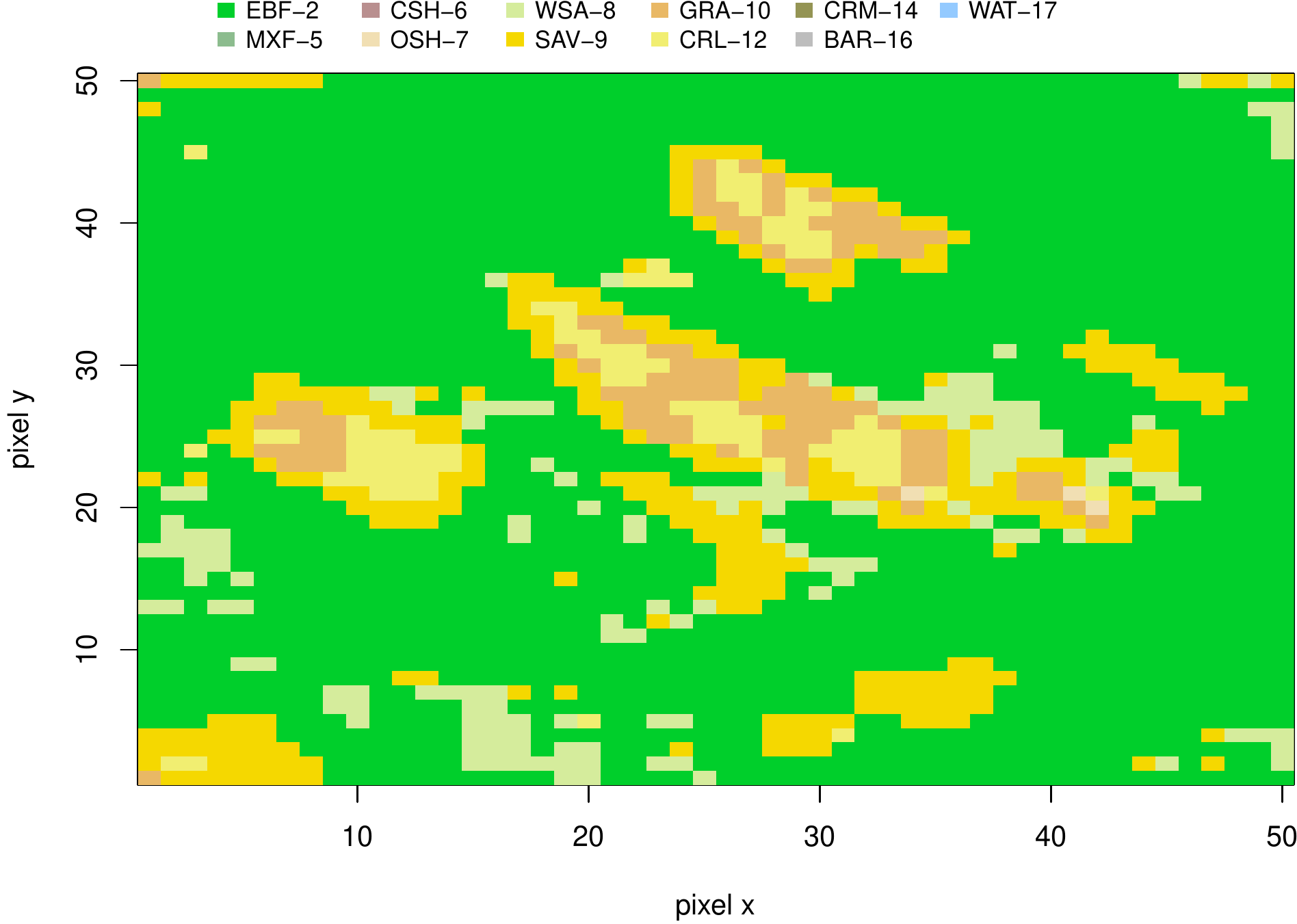}
\end{tabular}
\caption{Results of change point analysis in the Xingu River study region.
Top row: estimated change points, land cover class compositions (top
panels), and projected profiles (bottom panels) for two representative pixels
in the study region. In top panels: probability of first change point with
change weights given in color according to land cover class (see legend),
dashed line marks probability of no change, dark gray bar marks probability of
second change to background class (recovery), light gray background represents
deforestation percentages from ground-truth reference; in bottom panels:
PC-projected spectro-temporal data profiles, with hollow points marking
EM-imputed values, solid lines representing mean land cover profiles, and
dashed lines mean profiles for background class.
Bottom-left panel: probabilities of land cover changes (bars) given that
change has occurred; jittered points highlight the same probabilities but when
these probabilities are maximized for the respective change class.
Bottom-right panel: overall land cover classification based on inferred
change.}
\label{fig:xingu-50-pixels}
\end{figure}

An important feature of our model is to not only represent changes but to also
characterize these changes according to land cover classes. As an example, the
top row of Figure~\ref{fig:xingu-50-pixels} illustrates the results for two
typical pixels in the study region. In each panel, the top plot shows the
probability of change points, $\Pr(\rho_{1v} \given Y_v; \hat{\Theta})$,
further stratified by class probabilities $\Pr(W_v \given Y_v; \hat{\Theta})$
in a colored bar, $\Pr(\rho_{2v} \given Y_v; \hat{\Theta})$ in a dark gray bar
if positive, and the deforestation percentages from the ground-truth reference
in light gray in the background. The dashed line represents the probability of
no change. The bottom plot depicts PC-projected NBAR values as
in~\eqref{eq:pca} with $K=3$ with EM-imputed values in hollow points; the
solid lines in each year outline the projected mean profile for the inferred
class in the year, while dashed lines represent background (EBF) yearly
profiles. As we can see from the left panel, the changed class profile fits
the data reasonably well, and hence the high probability of change at year~7;
on the other hand, in the right panel the projected data does not seem to
follow class profiles closely and so the no-change probability is closer to
the now smaller change probability and the class to which the pixel changed is
less certain.

The bottom row portrays to which land cover classes pixels change (left panel,
posterior conditional probabilities $\Pr(W_v \given Y_v, \rho_{1v} > J)$)
and how these change classes are distributed spatially in the study region
(right panel.) In the left barplot, we see that the most common change classes
are, in order, savannas (IGBP class~9), woody savannas (IGBP~8), grasslands
(IGBP~10), and croplands (IGBP~12.) The jittered gray points highlight the
probability of changing to each class $C$ when
$\Pr(W_v \given Y_v, \rho_{1v} > J)$ is maximized at $C$. The right panel
displays the study region with each pixel colored by either the background EBF
class if there is no inferred change, or by the class that maximizes
$\Pr(W_v \given Y_v, \rho_{1v} > J)$ in case of change. Thus, the panel
contains the same spatial patterns as in the top two leftmost panels in
Figure~\ref{fig:xingu-50}, but it adds a characterization of change according
to land cover.

\section{Discussion}
As the simulation study in Section~\ref{sec:simstudy} indicates, the
proposed model and EM inferential routine yield better results than a
state-of-the-art alternative method. Our better performance can be explained
mainly by three factors: first, our proposed model incorporates data from all
bands, instead of relying on particular bands or combined statistics (e.g.\@
NDVI and EVI~\citep{myneni95,huete02}); missing data is ubiquitous in remote
sensing and while many methods depend on extraneous gap-filling procedures,
our method accommodates missing data consistently with our model via
expectation; finally, our model is more flexible since we allow for at most
two change points to capture recovery from change.

Our proposed methodology has also performed well in the real-world case study
in Section~\ref{sec:casestudy}. The results are in very good agreement with
the ground-truth reference. Interestingly, as we can see in
Figure~\ref{fig:xingu-50}, the inferred changes seem to follow a clear spatial
pattern usually going northwest to southeast and operating on clusters; this
effect is reassuring since the model makes no provisions for spatial
interactions and so the pattern is fortuitous. Concordance is generally high
over the whole study region with low concordance pixels being localized to
change cluster borders---which we attribute to pixels with mixed class
compositions due to transitions from background (see, for
example,~\citep{jin05,lunetta06})
---or to small clusters. These small clusters capture
larger discrepancies with the reference about the existence of change and/or
deforestation.

The two exemplar pixels in Figure~\ref{fig:xingu-50-pixels} highlight the
two major types of discrepancies to the ground-truth reference that lead to
lower concordance values in Figure~\ref{fig:xingu-50}. In the top left panel
we have a low deforestation percentage but high estimated probability of
change at year~7 to savanna; this pixel belongs to the small cluster in the
southeast corner of the region. Given that the spectro-temporal profile for
savanna is similar to the profile for evergreen broadleaf forests (EBF, the
background land cover class), and that savannas have from~10 to~30\% of
forest canopy cover, it is reasonable to confuse this land cover class with a
low deforestation profile. In the top right panel we summarize the results
from the EM method for a pixel in the southern border of the big change
cluster in the middle of the study region. In this case, the class
fragmentation at the change year, year~5, and possible recovery at year~8 can
be attributed to deforestation and/or degradation at sub-pixel scale.

As we can see in the bottom left panel of Figure~\ref{fig:xingu-50-pixels},
in the Xingu River region case study most land cover classes in the
estimated change segments are woody savannas, savannas, grasslands, or
croplands (IGBP classes~8, 9, 10, and~12, respectively). Croplands and
grasslands are often found in regions with earlier change points (darker
regions in the top left plot in Figure~\ref{fig:xingu-50}), and might
correspond to new land uses such as soy plantations and cattle ranching farms.
In contrast, later change point regions are often classified as woody
savannas, which have higher canopy density and might signal recent
deforestation. Savannas have lower canopy density and are localized to either
border pixels, as a transition land cover class, or to isolated islands; these
smaller stranded regions could correspond to degradation areas, a more veiled
form of deforestation. Interestingly, most discrepancies to the reference
deforestation percentages overlap with this land cover class; this can be
explained by lower deforestation percentages in these regions.

\section{Conclusion}
Detecting changes in land cover can provide crucial information for land use
policy, natural resource management, and ecosystem modeling efforts. Remote
sensing offers a spectrally and temporally rich source of data with which to
make inference about changes in land cover at broad spatial scales.
Unfortunately, missing values pervade most datasets for a multitude of
reasons.

In this article we proposed a hierarchical model for identifying
conversion-type changes in MODIS time series which accounts for missing data.
The collection of MODIS training sites for the IGBP classification scheme is
extensive and provides a useful resource for characterizing these
high-dimensional data. We use these training data to estimate model parameters
for 11 IGBP land cover classes including our background class: Evergreen
Broadleaf Forest. With these estimates in hand we proceed to analyze pixels
independently with an EM algorithm to detect the presence or lack of change
points. The change points we identify characterize distributional changes from
EBF to one of the other IGBP land cover classes present in our training
dataset.

Not only can our approach identify change points, but the posteriors
in~\eqref{eq:wvem} can be used to informally assess what class or classes the
changed data represent. The methodology we propose here has two distinctive
features: first, while our method is probably best used to find abrupt changes
in time series, such as disturbances, it is flexible enough to handle gradual
changes by suitably defining change probabilities $\pi_0$ and $\pi_R$ and
fitting class probabilities $\alpha$; moreover, the methodology we propose
allows for recovery from change. These two important features are essential
to characterizing and interpreting changes and are, in particular, essential
to remote sensing applications. We note that hyper-prior parameters $\pi_0$
and $\pi_R$ control how robust the method is to outliers and should be
carefully elicited based on similar study regions.

In general, our EM algorithm could be used successfully on data or land cover
displaying a conversion-type change. To accommodate other types of disturbances
such fire and logging, our model would require exemplars from these
situations. That is, we would need to characterize the surface after a fire or
after logging in the parameter estimates of~\eqref{eq:lik} (i.e.\ training
data for ``post-fire'' or ``post-logging'') in order to detect these kinds of
changes in new pixels. 

We demonstrated the effectiveness of our method with a simulation study and a
case study in the Xingu River Basin. Our results indicate that our method
performs better than state-of-the-art methods and has high concordance to
ground-truth references. More specifically, we recovered nicely the spatial
and temporal configuration of changes in the study regions and were able to
interpret the changes by their inferred land cover classes and spatial
localization. Overall, our method produced satisfying results and should be
considered for detecting conversion-type changes in remotely sensed time
series that contain missing data.

As future work we intend to extend this method to formally account for changes
in space, that is, not only in time, and to investigate an alternative
estimator for change configurations that maximizes the posterior expected
accuracy, that is, to define
$\hat{\rho}_A := \argmax_{\tilde{\rho}}
\Exp_{\rho \given Y} \big[A(\rho, \tilde{\rho})\big]$
and devise a computationally efficient method to obtain $\hat{\rho}_A$.


\section*{Acknowledgements}
Hunter Glanz was supported by funding from NASA under grant number NNX11AG40G.
Xiaoman Huang was supported by NASA grant numbers NNX11AE75G and NNX11AG40G.
Luis Carvalho was supported by NSF grant DMS-1107067. 


\appendix
\section{Expectation-Maximization Derivations}

To derive the EM updates in Section~\ref{sec:cpdem} we need
\[
\begin{split}
Q(\Theta, \Theta^{(t)})
&= \Exp_{Z, W \given Y; \Theta^{(t)}} \Big[ \log\Pr(\Theta, Z, W, Y) \Big] \\
&= \Exp_{Z, W \given Y; \Theta^{(t)}} \Bigg[
\sum_{v \in \mathscr{R}} \log\Pr(Z_v, Y_v \given W_v, \rho_v) +
\log\Pr(\rho_v) \\
&\qquad + I(\rho_{1v} < J)\log\Pr(W_v \given \boldsymbol{\alpha}) +
\log\Pr(\boldsymbol{\alpha}) \Bigg],
\end{split}
\]
as defined in~\eqref{eq:emupdate}. The indicator $I(\rho_{1v} < J)$ filters
pixels that have at least one change. To derive the conditional updates for
$\boldsymbol{\alpha}$ and $\rho_v$ for each $v \in \mathscr{R}$ we identify
two functions that capture the terms in $Q$ that depend on
$\boldsymbol{\alpha}$,
\begin{equation}
\label{eq:qalpha}
\begin{split}
Q_{\alpha}(\Theta, \Theta^{(t)}) &=
\Exp_{Z, W \given Y; \Theta^{(t)}}
\Bigg[ \sum_{v : \rho_{1v} < J}\log\Pr(W_v \given \boldsymbol{\alpha}) +
\log\Pr(\boldsymbol{\alpha}) \Bigg] \\
&= \sum_{v : \rho_{1v} < J} \Exp_{W_v \given Y_v; \Theta^{(t)}}
\Big[ \log\Pr(W_v \given \boldsymbol{\alpha}) \Big] +
\log\Pr(\boldsymbol{\alpha}),
\end{split}
\end{equation}
and on $\rho_v$ at pixel $v$,
\begin{equation}
\label{eq:qv}
\begin{split}
Q_{\rho, v}(\Theta, \Theta^{(t)})
& = \Exp_{Z_v, W_v \given Y_v; \Theta_v^{(t)}}
\Big[ \log\Pr(Z_v, Y_v \given W_v, \rho_v) + \log\Pr(\rho_v) \\
& \qquad + I(\rho_{1v} < J)\log\Pr(W_v \given \boldsymbol{\alpha}) \Big] \\
& = \Exp_{Z_v, W_v \given Y_v; \Theta_v^{(t)}}
[ \log\Pr(Z_v, Y_v \given W_v, \rho_v) ] + \log\Pr(\rho_v) \\
& \qquad + I(\rho_{1v} < J)\Exp_{Z_v, W_v \given Y_v; \Theta_v^{(t)}}
[\log\Pr(W_v \given \boldsymbol{\alpha}) ]. \\
\end{split}
\end{equation}
Note that $Q(\Theta, \Theta^{(t)}) = \sum_v Q_{\rho,v}(\Theta, \Theta^{(t)}) +
\log\Pr(\boldsymbol{\alpha})$ and that the term
$I(\rho_{1v} < J)\Pr(W_v \given \boldsymbol{\alpha})$ is shared between
$Q_{\alpha}$ and $Q_{\rho,v}$.

\subsection*{Updating $\boldsymbol{\alpha}$}
Let us start with the $\boldsymbol{\alpha}$-update in Step 2.a; we need to
optimize $Q$ with respect to $\boldsymbol{\alpha}$ subject to the constraint
$h(\boldsymbol{\alpha}) = \sum_{g \in \mathscr{C}} \alpha_g - 1 = 0$. To this
end, we define a Lagrange multiplier $\lambda$ and solve
\begin{multline*}
\frac{\partial}{\partial \alpha_k}\Big[
Q_{\alpha}(\Theta, \Theta^{(t)}) - \lambda h(\boldsymbol{\alpha})
\Big] 
= \frac{\partial}{\partial \alpha_k}\Bigg[
\sum_{v : \rho_{1v} < J} \Exp_{W_v \given Y_v; \Theta^{(t)}} \Bigg[
\sum_{g \in \mathscr{C}} I(W_v = g) \log \alpha_g \Bigg] \\
+ \sum_{g \in \mathscr{C}} (\pi_g - 1) \log \alpha_g
- \lambda \sum_{g \in \mathscr{C}} \alpha_g
\Bigg] = 0,
\end{multline*}
and so, fixing $\rho_v$ to its value in the previous iteration, $\rho_v^{(t)}$, we
get the update in~\eqref{eq:alphaem},
\[
\begin{split}
\alpha_k^{(t+1)} &= \frac{\pi_k - 1 +
\sum_{v : \rho_{1v}^{(t)} < J} \Exp_{W_v \given Y_v; \Theta^{(t)}} [ I(W_v = k) ]}
{\lambda} \\
&= \frac{\sum_{v : \rho_{1v}^{(t)} < J} \Pr(W_v = k \given Y_v; \Theta^{(t)})
+ \pi_k - 1}
{\sum_{g \in \mathscr{C}} \big[\sum_{v : \rho_{1v}^{(t)} < J}
\Pr(W_v = g \given Y_v; \Theta^{(t)}) + \pi_g - 1\big]} \\
&= \frac{\sum_{v : \rho_{1v}^{(t)} < J} \Pr(W_v = k \given Y_v; \Theta^{(t)}) + 
\pi_k - 1}
{N_v^{(t)} + \sum_{g \in \mathscr{C}} \pi_g - |\mathscr{C}|},
\end{split}
\]
where $N_v^{(t)} = \sum_{v : \rho_{1v}^{(t)} < J}
\sum_{g \in \mathscr{C}} \Pr(W_v = g \given Y_v; \Theta^{(t)})
= \sum_{v : \rho_{1v}^{(t)} < J} 1$ is the number of pixels with changes. To
compute the update we just need the expression in~\eqref{eq:wvem},
for $k \in \mathscr{C}$,
\[
\begin{split}
\Pr(W_v = k \given Y_v; \Theta^{(t)}) &=
\frac{\Pr(Y_v \given W_v = k; \Theta^{(t)}) \Pr(W_v = k; \Theta^{(t)})}
{\sum_{g \in \mathscr{C}}
\Pr(Y_v \given W_v = g; \Theta^{(t)}) \Pr(W_v = g; \Theta^{(t)})} \\
&= \frac{\alpha_k^{(t)} \Pr(Y_v \given W_v = k; \Theta^{(t)})}
{\sum_{g \in \mathscr{C}} \alpha_g^{(t)} \Pr(Y_v \given W_v = g;
\Theta^{(t)})}.
\end{split}
\]

\subsection*{Updating $\rho_v$}
In Step 2.b we fix $\boldsymbol{\alpha}$ and update the remaining parameters
in $\Theta$. We update them jointly, but in parallel for each pixel.
The last term in $Q_{\rho,v}$ is already known from the last section, and we
condition $\boldsymbol{\alpha}$ to its recently updated value
$\boldsymbol{\alpha}^{(t+1)}$:
\begin{equation}
\label{eq:wvapp}
\Exp_{Z_v, W_v \given Y_v; \Theta_v^{(t)}}
\Big[ \log\Pr(W_v \given \boldsymbol{\alpha}) \Big] =
\sum_{g \in \mathscr{C}} \Pr(W_v = g \given Y_v; \Theta^{(t)})
\log \alpha_g^{(t+1)}.
\end{equation}

Now we just need to obtain
\begin{multline}
\label{eq:xvapp}
\Exp_{Z_v, W_v \given Y_v; \Theta_v^{(t)}}
\Big[ \log\Pr(X_v \given W_v, \rho_v) \Big]
= \Exp_{W_v \given Y_v; \Theta_v^{(t)}} \Bigg[
\sum_{g \in \mathscr{C}} I(W_v = g) \\
\Exp_{Z_v \given W_v = g, Y_v; \Theta_v^{(t)}}
\Big[ \log\Pr(X_v \given W_v, \rho_v) \Big]
\Bigg] \\
= -\frac{1}{2} \Bigg(
\sum_{i \in \text{BG}(\rho_v)} S(X_{iv}; \mu_F, \tilde{\Sigma}_F) + 
\sum_{i \not\in \text{BG}(\rho_v)} \sum_{g \in \mathscr{C}} 
\Pr(W_v = g \given Y_v; \Theta^{(t)}) S(X_{iv}; \mu_g, \tilde{\Sigma}_g) \Bigg),
\end{multline}
where $\tilde{\Sigma}_F = \Sigma_F + \kappa_0 I_{BT}$, $\tilde{\Sigma}_g =
\Sigma_g + \kappa_c I_{BT}$ for $g \in \mathscr{C}$, as in the main text, and,
if $X \sim N(\mu, \Sigma)$ with missing entries at indices $\miss$,
\[
S(X; \mu, \Sigma) := \log|\Sigma| + \Exp_{X_{\miss} \given X_{-\miss}}\Big[
{(X - \mu)}^\top \Sigma^{-1} (X - \mu) \Big].
\]

To evaluate $S$, we need
\[
\begin{split}
\Exp_{X_{\miss} \given X_{-\miss}}\Big[
{(X - \mu)}^\top \Sigma^{-1} (X - \mu) \Big]
&= \Exp_{X_{\miss} \given X_{-\miss}}\Big[
\text{tr}\Big\{{(X - \mu)}^\top \Sigma^{-1} (X - \mu) \Big\}
\Big] \\
&= \text{tr}\Big\{\Sigma^{-1} 
\Exp_{X_{\miss} \given X_{-\miss}}\Big[(X - \mu){(X - \mu)}^\top\Big]\Big\} \\
&= {(\tilde{X} - \mu)}^\top \Sigma^{-1} (\tilde{X} - \mu) \\
& \qquad +
\text{tr}\Big\{\Sigma^{-1} \Var_{X_{\miss} \given X_{-\miss}}[X] \Big\},
\end{split}
\]
since, with $\tilde{X} = \Exp_{X_{\miss} \given X_{-\miss}}[X]$, we have the
Pythagorean relationship
\[
\begin{split}
\Exp_{X_{\miss} \given X_{-\miss}}\Big[(X - \mu){(X - \mu)}^\top\Big]
&= \Exp_{X_{\miss} \given X_{-\miss}}\Big[
(X - \tilde{X}){(X - \tilde{X})}^\top \\
& \qquad + (\tilde{X} - \mu){(\tilde{X} - \mu)}^\top \Big] \\
&= \Var_{X_{\miss} \given X_{-\miss}}[X] + 
(\tilde{X} - \mu){(\tilde{X} - \mu)}^\top.
\end{split}
\]
Let us denote by $V(X; \Sigma) := \Var_{X_{\miss} \given X_{-\miss}}[X]$.
Clearly, $\tilde{X}_{-\miss} = X_{-\miss}$ and so ${V(X; \Sigma)}_{jk} = 0$
wherever $j \not\in \miss$ or $k \not\in \miss$. The remaining entries in
$\tilde{X}$ and $V(X; \Sigma)$ are known from
\begin{multline*}
X_{\miss} \given X_{-\miss} \sim N\Big(
\mu_{\miss} + {(\Sigma_{-\miss, \miss})}^\top {(\Sigma_{-\miss, -\miss})}^{-1}
(X_{-\miss} - \mu_{-\miss}), \\
\Sigma_{\miss, \miss} -
{(\Sigma_{-\miss, \miss})}^\top {(\Sigma_{-\miss, -\miss})}^{-1}
\Sigma_{-\miss, \miss}
\Big).
\end{multline*}
Thus,
\[
S(X; \mu, \Sigma) = \log|\Sigma| + 
{(\tilde{X} - \mu)}^\top \Sigma^{-1} (\tilde{X} - \mu) +
\sum_{j,k \in \miss} {(\Sigma^{-1})}_{jk} {V(X; \Sigma)}_{jk},
\]
which yields the definition in~\eqref{eq:Sem}.

Finally, putting together~\eqref{eq:wvapp} and~\eqref{eq:xvapp} in the
definition of $Q_{\rho,v}$, and since $\argmax_{\rho} Q_{\rho,v} =
\argmin_{\rho} -2Q_{\rho,v}$, we have the update expression
in~\eqref{eq:rhoem}.


\bibliographystyle{chicago}
\bibliography{biblio}

\end{document}